\documentclass[aps,prb,showpacs,twocolumn]{revtex4-1}
\usepackage[hypertex]{hyperref}
\usepackage{amsmath,amsfonts,bm,graphicx,amssymb,color}

\newcommand{\llangle}{\langle\!\langle}
\newcommand{\rrangle}{\rangle\!\rangle}
\newcommand{\GBequal}{\underset{\mathrm{binomial}}{\overset{\mathrm{generalized}}{=}}}
\newcommand{\GBrightarrow}{\underset{\mathrm{binomial}}{\overset{\mathrm{generalized}}{\rightarrow}}}
\newcommand{\hatH}{\hat{H}}

\begin{document}

\title{Factorial cumulants reveal interactions in counting statistics}
\author{Dania Kambly}
\author{Christian Flindt}
\author{Markus B\"uttiker}
\affiliation{D\'epartement de Physique Th\'eorique, Universit\'e de Gen\`eve, CH-1211 Gen\`eve, Switzerland}
\date{\today}

\begin{abstract}
Full counting statistics concerns the stochastic transport of electrons in mesoscopic structures. Recently it has been shown that the charge transport statistics for noninteracting electrons in a two-terminal system is always generalized binomial: it can be decomposed into independent single-particle events, and the zeros of the generating function are real and negative. Here we investigate how the zeros of the generating function move into the complex plane due to interactions and demonstrate that the positions of the zeros can be detected using high-order factorial cumulants. As an illustrative example we consider electron transport through a Coulomb blockade quantum dot for which we show that the interactions on the quantum dot are clearly visible in the high-order factorial cumulants. Our findings are important for understanding the influence of interactions on counting statistics, and the characterization in terms of zeros of the generating function provides us with a simple interpretation of recent experiments, where high-order statistics have been measured.
\end{abstract}



\maketitle

\section{Introduction} \label{sec:intro}

\begin{figure}
\begin{center}
\includegraphics[width=0.4\textwidth, trim = 0 0 0 0, clip]{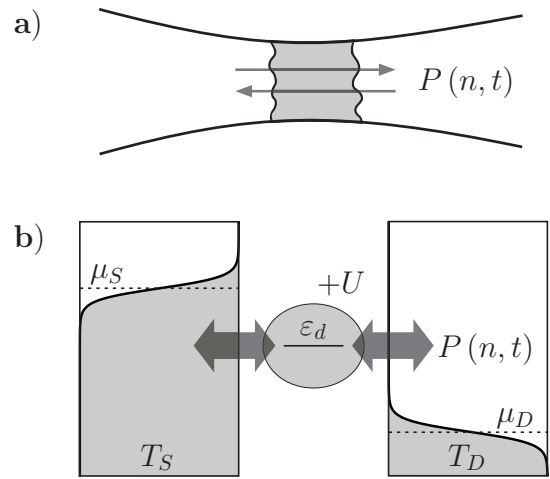}
\caption{Nanoscale conductors. (a) Generic transport setup consisting of a nanoscale conductor (gray) connected to source and drain electrodes. Electron flow in both directions (arrows) is allowed. The probability distribution for the number of electrons $n$ that have been collected in the drain electrode during the time span $[0,t]$ is denoted $P(n,t)$. (b) Single-level quantum dot coupled via tunnel barriers to source $(S)$ and drain $(D)$ electrodes, kept at temperatures $T_S$ and $T_D$, respectively, and chemical potentials $\mu_S$ and $\mu_D$. The single-level energy of the quantum dot is $\varepsilon_d$ and the on-site Coulomb interaction is $U$ (see Sec. \ref{sec:model}).}
\label{Fig:setup}
\end{center}
\end{figure}

Full counting statistics (FCS) has been a topic of active research for nearly two decades.\cite{Levitov1993,Nazarov2003} FCS describes the statistics of charge transport through mesoscopic conductors [Fig.\ \ref{Fig:setup}(a)] and is expected to provide more information about the physical processes inside a conductor compared to what is available from the mean current and the shot noise only. Within this framework it is natural to ask how such additional information can be extracted from the high-order statistics and what quantities or measures are most suitable to this end. This core problem of FCS constitutes the main focus of the present work.

On the experimental side, FCS has recently gained considerable impetus due to a number of measurements of high-order statistics in nanoscale systems. While earlier experiments were restricted to the first few moments or cumulants of the current, high-order cumulants of the charge transport are now becoming experimentally accessible. The fourth and fifth cumulants of the current have recently been measured both in Coulomb blockade quantum dots\cite{Gustavsson2009} and in avalanche diodes.\cite{Gabelli2009} Quantum dots, in particular, have emerged as useful sources of high-accuracy counting statistics. In these systems, the low (kilo-hertz) tunneling rates allow for detection of individual electrons in real-time using a nearby quantum point contact whose conductance is sensitive to the charge occupations of the dots.\cite{Gustavsson2006,Sukhorukov2007,Gustavsson2009} Although quantum effects are typically suppressed at the long time scales characterizing the charge transport, quantum dots provide a unique setting to experimentally test theoretical predictions for high-order statistics. Remarkably, time-dependent cumulants of the transferred charge beyond the 15th order have recently been measured in single-electron transport through a Coulomb blockade quantum dot\cite{Flindt2009,Fricke2010a,Fricke2010b} and a wealth of statistical data is now available.

On the theory side, several techniques have been developed for calculating the statistics of transferred charges. In the seminal work by Levitov and Lesovik,\cite{Levitov1993} the FCS of noninteracting electrons propagating coherently through a conductor was expressed by a determinant formula containing the scattering matrix of the problem.\cite{Levitov1993,Ivanov1993,Levitov1996,Nazarov2002} In many-channel conductors the statistics is predominantly classical and can be described using a Langevin--Boltzmann equation.\cite{Nagaev2002,Nagaev2010} A powerful and elegant formulation in this semiclassical regime is the stochastic path integral approach, which was introduced for FCS in the pioneering works by Pilgram \emph{et al.}\cite{Pilgram2003,Jordan2004,Nagaev2004} For interacting systems, the FCS can be related to a generalized master equation describing the charge transport\cite{Bagrets2003,Emary2007,Flindt2008,Flindt2010} or obtained using Keldysh Greens functions.\cite{Nazarov1999,Nazarov2002}

The central object in the theory of FCS is the generating function (GF) for the probability distribution $P(n,t)$ of the number of transferred electrons $n$. Its analytic properties as well as its symmetries as a function of voltage, temperature, and counting fields, which, for example, lead to fluctuation relations,\cite{Harris2007,Esposito2009,Foerster2008,Sanchez2010} are therefore of fundamental interest. If the transport process consists of independent, elementary events, the GF may be factorized according to these. This has been pointed out by Vanevi\'c, Nazarov and Belzig, who identified the elementary transport processes in a quantum conductor driven by a time-dependent voltage.\cite{Vanevic2007,Vanevic2008} Importantly, Abanov and Ivanov have shown recently that for  noninteracting electrons in a two-terminal setup, the GF can always be factorized into single-particle events and the zeros of the GF correspondingly lie on the negative real axis.\cite{Abanov2008,Abanov2009} This form of the counting statistics has been dubbed generalized binomial by Hassler \emph{et al.}\cite{Hassler2008} Interactions, however, can change these properties and cause the zeros to move into the complex plane.\cite{Abanov2009} As evident from this discussion, the zeros of the GF are of crucial importance for understanding the physical mechanisms that determine the transport statistics. However, given a complicated GF or even actual experimental data, it is not clear how one can tell, whether or not the statistics are generalized binomial and if the charges interacted inside the conductor.

Traditionally within FCS, the probability distribution $P(n,t)$ has been characterized by its cumulants, from which one could hope to extract information about the physical processes inside the conductor.\cite{Nazarov2003} The cumulants are related to derivatives of the GF. Several theoretical studies, however, have found that the high-order cumulants tend to oscillate strongly as some system parameter is varied.\cite{Pilgram2003b,Forster2005,Forster2007,Flindt2008,Novaes2008,Golubev2010,Dominguez2010} Recently, this phenomenon has been explained on general grounds and it has been shown that the high-order cumulants for almost any non-Gaussian distribution should oscillate as functions of basically any system parameter,\cite{Flindt2009} following work by Berry on high derivatives of smooth functions.\cite{Berry2005} The fact that these oscillations are so generic could indicate that the high-order cumulants are fragile objects from which it is difficult to extract information about a particular system: The information in the high-order cumulants seems to be masked by the oscillatory behavior that should be present in almost any system. However, as we explain below, such oscillations do not prevent high-order statistics from bearing information.

It is the purpose of this paper to argue that factorial cumulants provide an alternative characterization of the probability distribution $P(n,t)$ that is particularly useful for describing the high-order statistics of nanoscale transport. In mesoscopic physics, factorial cumulants have so far received only limited attention, except in studies of photons emitted from a quantum point contact,\cite{Beenakker2004} and, very recently, in connection with current fluctuations and entanglement entropy.\cite{Song2010} However, as we show, the high-order factorial cumulants provide a simple description of generalized binomial statistics and they directly reflect the zeros of the GF. In fact, for noninteracting systems, the high-order factorial cumulants do not oscillate (unlike the ordinary cumulants), no matter what parameter is varied. In contrast, if a high-order factorial cumulant oscillates as a function of some parameter, it cannot be describing generalized binomial statistics and the electrons must have interacted inside the conductor. We illustrate these points with a model of transport through a quantum dot [Fig.\ \ref{Fig:setup}b], for which we show how interactions on the dot cause the zeros of the GF to move into the complex plane as clearly seen in the factorial cumulants.

The paper is organized as follows: In Sec. \ref{sec:counting_statistics} we introduce the general framework of FCS, including the GF and the moments and cumulants, as well as their factorial counterparts. We show that the factorial cumulants are particularly simple for generalized binomial statistics, and we give an asymptotic expression for the high-order factorial cumulants, which is directly related to the zeros of the GF. We show that the behavior of the high-order factorial cumulants changes drastically as the zeros move into the complex plane due to interactions. In Sec. \ref{secIII} we illustrate our findings using a model of electron transport through a quantum dot, which is weakly coupled to source and drain electrodes. At low bias voltages, the quantum dot can only be empty or singly occupied. This corresponds to the recent experiment by Fricke \emph{et al.},\cite{Fricke2010b} and we show how the measured oscillations of the 15th cumulant as a function of time and coupling to the leads can be explained by the motion of the zeros of the GF. In contrast, the factorial cumulants do not oscillate. As the voltage difference between the leads is increased, the quantum dot can be occupied also by two electrons at a time, and the on-site Coulomb interaction now strongly affects the FCS. The zeros of the GF move into the complex plane, which is clearly visible in the high-order factorial cumulants. Finally, Sec. \ref{sec:conclusions} contains our concluding remarks as well as a number of open questions and directions for future research. Appendixes \ref{app:finite_time_fcs} and \ref{app:B}, respectively, describe methods for calculating high-order (factorial) moments and cumulants at finite times from a master equation and for extracting the position of the zeros from the high-order factorial cumulants.

\section{Full counting statistics}
\label{sec:counting_statistics}

We consider a generic transport setup in which a nanoscale conductor is connected to source and drain electrodes [Fig.\ \ref{Fig:setup}a]. A bias voltage (possibly time dependent) between the electrodes drives electrons through the conductor, and we denote by $P(n,t)$ the probability distribution of the number of electrons $n$ that have traversed the nanoscale conductor during the time span $[0,t]$. Depending on the direction of the tunneling events between the source and the drain electrodes, the number of transferred electrons $n$ can be either positive or negative. For concreteness we define electron flow into the drain electrode as the positive direction of the current. Examples of nanoscale conductors used in recent counting statistics experiments include tunnel junctions,\cite{Reulet2003,Bomze2005} quantum point contacts,\cite{Gershon2008} double quantum dots,\cite{Fujisawa2006} and nanowires,\cite{Choi2009} but for the discussion in this section it is not necessary to specify the details of the system.

\subsection{Generating function}

The object of main concern in FCS is the GF, defined as
\begin{equation}
\mathcal{G}(z,t)=\sum_{n}P(n,t)z^n.
\label{eq:GF}
\end{equation}
The GF encodes the full information about the probabilities $P(n,t)$ and it allows us to introduce (factorial) moments and (factorial) cumulants in the following. As mentioned in Sec. \ref{sec:intro}, calculations of the GF can be approached with several different techniques, depending on the specific system at hand. Interestingly, however, Abanov and Ivanov have recently shown that the counting statistics for a two-terminal scattering problem involving noninteracting electrons can always be decomposed into independent single-particle events for any form of the scattering matrix and at any temperature.\cite{Abanov2008,Abanov2009} Formally, this can be expressed as a factorization of the GF of the form\cite{Abanov2008,Abanov2009}
\begin{equation}
\mathcal{G}(z,t)\GBequal z^{-Q}\prod_{i}\mathcal{G}_i(z,t),
\label{eq:GB_GF}
\end{equation}
where $\mathcal{G}_i(z,t)=1-p_i+p_iz$ is the binomial GF corresponding to a single-particle event occurring with probability $0\leq p_i\leq 1$, depending on time $t$ as well as all other parameters of the system. The factor $z^{-Q}$ corresponds to a deterministic background charge transfer $Q=\sum_ip_i-\langle n\rangle\geq 0$ opposite to the positive direction, where $\langle n\rangle$ is the mean value of the total transferred charge. For unidirectional transport, we have $Q=0$.

A statistical distribution given by Eq.\ (\ref{eq:GB_GF}) has been dubbed generalized binomial statistics.\cite{Hassler2008} Importantly, the result of Abanov and Ivanov implies that if a GF cannot be factorized as in Eq.\ (\ref{eq:GB_GF}) with real probabilities $p_i$, it cannot be describing noninteracting electrons. We note, however, that the opposite is not true: even in the presence of interactions, the statistics may still be generalized binomial.\cite{Abanov2009}

\subsection{Moments and cumulants}

We now turn to the moments and cumulants of $P(n,t)$, which are commonly used to characterize the probability distribution. The moments $\langle n^m\rangle$ can be found from the moment generating function which is obtained from the GF [Eq.\ (\ref{eq:GF})] via the substitution $z\rightarrow e^{z}$,
\begin{equation}
\mathcal{M}(z,t)= \mathcal{G}(e^{z},t)=\sum_{n}P(n,t)e^{nz}.
\label{eq:MGF}
\end{equation}
The moments of the transferred charge are given by the derivatives of the moment generating function with respect to $z$, evaluated at $z=0$,
\begin{equation}
\langle n^m\rangle(t) =\sum_{n}n^mP(n,t) = \left.\partial^m_{z}\mathcal{M}(z,t)\right|_{z\rightarrow 0}.
\label{eq:moments}
\end{equation}
The cumulant generating function (CGF) is defined as
\begin{equation}
\mathcal{S}(z,t)= \ln\left[\mathcal{M}(z,t)\right]=\ln\left[\mathcal{G}(e^{z},t)\right]
\label{eq:CGF}
\end{equation}
and the cumulants are similarly defined as derivatives of the CGF at $z=0$:
\begin{equation}
\llangle n^m\rrangle(t) = \left.\partial^m_{z}\mathcal{S}(z,t)\right|_{z\rightarrow 0}.
\label{eq:cumulants}
\end{equation}
The first cumulant $\llangle n\rrangle=\langle n\rangle$ is the mean of $n$, the second $\llangle n^2\rrangle=\langle n^2\rangle-\langle n\rangle^2$ is the variance, and the third  $\llangle n^3\rrangle=\langle (n-\langle n\rangle)^3\rangle$ is the skewness. For a Poisson distribution all cumulants are equal to the mean $\llangle n^m\rrangle=\langle n\rangle$, while only the first and second cumulants are nonzero for a Gauss distribution, that is, $\llangle n^m\rrangle=0$ for $m>2$.

\subsection{Factorial moments and factorial cumulants}

A complementary characterization of the probability distribution is provided by the factorial moments and the corresponding factorial cumulants.\cite{Kendall1977,Johnson2005} The factorial moments are defined as follows
\begin{equation}
\langle n^m\rangle_F=\langle n\left(n-1\right)\cdots \left(n-m+1\right)\rangle.
\label{eq:moments2factorialmoments}
\end{equation}
It is easy to show that they are generated by the function
\begin{equation}
\mathcal{M}_F(z,t)= \mathcal{G}(z+1,t)=\sum_{n}P(n,t)(z+1)^{n},
\label{eq:FMGF}
\end{equation}
obtained from the GF [Eq.\ (\ref{eq:GF})] via the substitution $z\rightarrow z+1$. Thus
\begin{equation}
\langle n^m\rangle_F(t) = \left.\partial^m_z\mathcal{M}_F(z,t)\right|_{z\rightarrow 0},
\label{eq:FM}
\end{equation}
and similarly to the cumulants, the factorial cumulant generating function (FCGF) is defined as
\begin{equation}
\mathcal{S}_F(z,t)= \ln\left[\mathcal{M}_F(z,t)\right]=\ln\left[\mathcal{G}(z+1,t)\right],
\label{eq:FCGF}
\end{equation}
whose derivatives at $z=0$ deliver the factorial cumulants
\begin{equation}
\llangle n^m\rrangle_F(t) = \left.\partial^m_{z}\mathcal{S}_F(z,t)\right|_{z\rightarrow 0}.
\label{eq:fact_cumulants}
\end{equation}
The first two factorial cumulants are $\llangle n\rrangle_F=\langle n\rangle$ and $\llangle n^2\rrangle_F=\langle n^2\rangle-\langle n\rangle^2-\langle n\rangle$. For a Poisson distribution, only the first factorial cumulant is nonzero and $\llangle n^m\rrangle_F=0$ for $m>1$.

The factorial cumulants can be expressed in terms of the ordinary cumulants via the relations
\begin{equation}
\begin{split}
  \llangle n^1 \rrangle_F &= \llangle n^1 \rrangle, \\
  \llangle n^2 \rrangle_F &= \llangle n^2 \rrangle - \llangle n^1 \rrangle, \\
  \llangle n^3 \rrangle_F &= \llangle n^3 \rrangle - 3\llangle n^2 \rrangle + 2\llangle n^1 \rrangle, \\
\end{split}
\end{equation}
which, for arbitrary order $m$, read\cite{Johnson2005}
\begin{equation}
\llangle n^m \rrangle_F = \sum_{j=1}^m s\left(m,j\right)\llangle n^j \rrangle,
\end{equation}
where $s\left(m,j\right)$ are the Stirling numbers of the first kind. They can be generated from the relation $\left[\ln\left(1+x\right)\right]^j=j! \sum_{m=j}^{\infty} \left[s\left(m,j\right)x^m /m!\right]$.\cite{Johnson2005}

In the case of generalized binomial statistics [Eq.\ (\ref{eq:GB_GF})] the expression for the factorial cumulants becomes particularly simple:
\begin{equation}
\llangle n^m\rrangle_F \GBequal (-1)^{m-1}(m-1)!\left[\sum_i p_i^m-Q\right].
\label{eq:GB_FCs}
\end{equation}
This expression provides us with a direct test of whether or not a statistical distribution can be factorized into independent single-particle events as described by Eq.\ (\ref{eq:GB_GF}). In particular, for unidirectional transport (where $Q=0$), the factorial cumulants must have alternating signs as functions of the cumulant order $m$ due to the factor $(-1)^{m-1}$, if the statistics is generalized binomial. We remark that it is straightforward to extend this analysis to bi-directional transport. In that case, the quantity $\llangle n^m\rrangle_F-(-1)^{m-1}(m-1)!\langle n\rangle = (-1)^{m-1}(m-1)!\sum_i\left(p_i^m-p_i\right)$ has alternating sign as a function of $m$ for generalized binomial statistics. In this work we use Eq.\ (\ref{eq:GB_FCs}) to test whether the counting statistics of charge transport through a nanoscale conductor is generalized binomial. Importantly, factorial cumulants are measurable and the test is consequently immediately applicable to experimental data.

\subsection{High-order (factorial) cumulants}

As the zeros of the GF move into the complex plane due to interactions, the behavior of the high-order factorial cumulants changes compared to that of generalized binomial statistics [Eq.\ (\ref{eq:GB_FCs})]. To understand this, we note that both the CGF and the FCGF, defined in Eqs.\ (\ref{eq:CGF}) and (\ref{eq:FCGF}), respectively, have logarithmic singularities corresponding to the zeros of the GF. These singularities determine the high-order asymptotics of the ordinary and the factorial cumulants, respectively.\cite{Dingle1973,Berry2005,Flindt2009,Flindt2010} In this work we consider cases for which the (F)CGF has only logarithmic singularities, as is typical at finite times, but in the long-time limit it may have branch-point singularities, as we discuss at the end of this section.

We first note that the CGF (or the FCGF) close to a logarithmic singularity $z_j$ with degeneracy $\alpha_j$ behaves as
\begin{equation}
\mathcal{S}_{(F)}(z,t)\simeq \alpha_j\ln(z_j-z), \quad \text{$z$ close to $z_j$},
\label{eq:singLog_approx}
\end{equation}
with corresponding derivatives reading
\begin{equation}
\partial^m_{z}\mathcal{S}_{(F)}(z,t)\simeq -\alpha_j \frac{ (m-1)!}{(z_j-z)^m}, \quad \text{$z$ close to $z_j$}.
\label{eq:singLog_approx_der}
\end{equation}
The high-order derivatives evaluated at $z=0$, that is, the high-order (factorial) cumulants, can then be approximated as a sum over all singularities
\begin{equation}
\begin{split}
\llangle n^m \rrangle_{(F)} &= \partial_z^m\mathcal{S}_{(F)}(z,t)|_{z\rightarrow 0}\\
&\simeq  -\left(m-1\right)!\sum_j\alpha_j\frac{e^{-im\arg\left[z_j\right]}}{\left|z_j\right|^m}
\end{split}
\label{eq:high_cumu}
\end{equation}
according to the first Darboux approximation.\cite{Dingle1973,Berry2005}

Equation (\ref{eq:high_cumu}) shows that the high-order (factorial) cumulants are determined by the singularities closest to $z=0$, which dominate the sum for large $m$. Relative contributions from other terms are suppressed with the relative distance to $z=0$ and the power $m$.\cite{Flindt2009} If the closest (nondegenerate) singularity, denoted $z_0$, lies on the negative real axis such that $z_0=|z_0|e^{i\pi}$, Eq. (\ref{eq:high_cumu}) becomes particularly simple for large $m$ and reduces to
\begin{equation}
\llangle n^m \rrangle_{(F)} \rightarrow (-1)^{(m-1)}(m-1)!/\left|z_0\right|^{m}.
\label{eq:high_cumu_onesing}
\end{equation}
This is the case for high-order factorial cumulants corresponding to generalized binomial statistics [Eq.\ (\ref{eq:GB_GF})]. For unidirectional transport ($Q=0$) the corresponding FCGF is
\begin{equation}
\mathcal{S}_F(z,t)\GBequal \sum_i\ln{(1+p_i z)},
\end{equation}
which has logarithmic singularities at $z_j=-1/p_j\leq-1$. The high-order factorial cumulants are then dominated by the singularity $z_{\mathrm{max}}=-1/p_{\mathrm{max}}$ closest to $z=0$, where $p_{\mathrm{max}}$ is the largest probability among the $p_i$'s. We thus find
\begin{equation}
\llangle n^m\rrangle_F \GBrightarrow (-1)^{m-1}(m-1)! p_{\mathrm{max}}^m
\label{eq:FC_GB}
\end{equation}
for large $m$. For the above example, this conclusion could also have been reached directly from Eq.\ (\ref{eq:GB_FCs}) (for $Q=0$).

The singularities, however, do not always lie on the negative real axis, but in general they come in complex-conjugate pairs, ensuring that the (factorial) cumulants are real. In case only a single complex-conjugate pair of logarithmic singularities, $z_0$ and $z_0^*$,  is closest to $z=0$, Eq. (\ref{eq:high_cumu}) simplifies to
\begin{equation}
\llangle n^m \rrangle_{(F)} \rightarrow -\frac{2(m-1)!}{\left|z_0\right|^{m}}\cos\left(m\arg\left[z_0\right]\right)
\label{eq:high_cumu_twosing}
\end{equation}
for large $m$. In the case where $\arg\left[z_0\right]=\pi$, this reduces to the right-hand side of Eq.\ (\ref{eq:high_cumu_onesing}) multiplied by 2, since the two singularities are then degenerate.

Interestingly, our analysis shows that high-order factorial cumulants corresponding to generalized binomial statistics have a sign that is determined solely by the order $m$ via the factor $(-1)^{(m-1)}$. In contrast, if the statistics is not generalized binomial, the factorial cumulant of a given order $m$ will oscillate as a function of any parameter that changes the position of the singularities due to the factor $\cos\left(m\arg\left[z_0\right]\right)$, which also causes trigonometric oscillations as function of the order $m$.

In this section, we have analyzed the situation where the (F)CGF has logarithmic singularities due to zeros of the GF. It is, however, well known that the (F)CGF can have, for example, branch-point singularities in the long-time limit, and a generalization of the above analysis is necessary to treat such cases. The overall conclusions, however, remain intact also for branch-point singularities, and since we mainly consider finite times in this work, we do not encounter such situations. Instead, we refer the interested reader to Refs.\ \onlinecite{Dingle1973,Berry2005,Flindt2009} and, in particular, section IV of Ref.\ \onlinecite{Flindt2010} for a more general analysis of high-order derivatives and (factorial) cumulants for (F)CGFs with branch-point singularities.

In the next section, we illustrate how the statistics of charge transport through a quantum dot due to interactions can change from being generalized binomial, with factorial cumulants given by Eqs.\ (\ref{eq:GB_FCs}), to a different statistical distribution, with high-order factorial cumulants governed by Eq.\ (\ref{eq:high_cumu_twosing}).

\section{Coulomb blockade quantum dot}
\label{secIII}
\subsection{Model}
\label{sec:model}
We consider electron transport through a quantum dot (QD) with a single spin-degenerate level coupled to source and drain electrodes [Fig.\ \ref{Fig:setup}b]. The energy of the level is denoted by $\varepsilon_d$ and $U$ is the on-site Coulomb interaction. The Hamiltonian for the coupled system reads
\begin{equation}
\hatH=\hatH_{d}+\hatH_{T}+\hatH_R,
\label{eq:H}
\end{equation}
where
\begin{equation}
\hatH_d=\varepsilon_d(\hat{n}_\uparrow+\hat{n}_\downarrow)+U \hat{n}_\uparrow\hat{n}_\downarrow
\label{eq:H_d}
\end{equation}
is the Hamiltonian of the QD, tunneling between the QD and the leads is given by the term
\begin{equation}
\hatH_T=\sum_{\substack{k,\sigma,\\ \alpha=S,D}}\left(t_{\alpha k}~ \hat{c}_{\alpha k \sigma}^{\dag}\hat{d}_{\sigma}^{\phantom{\dag}}\quad + \quad \mathrm{h.\ c.\ } \right)
\label{eq:H_T}
\end{equation}
and the source ($\alpha=S$) and drain ($\alpha=D$) electrodes are described as reservoirs of free electrons with energy $\varepsilon_{\alpha k \sigma}$
\begin{equation}
\hatH_{R}=\sum_{\substack{k,\sigma,\\ \alpha=S,D}}\varepsilon_{\alpha k \sigma}~\hat{c}_{\alpha k \sigma}^{\dag}\hat{c}_{\alpha k \sigma}^{\phantom{\dag}}.
\label{eq:H_R}
\end{equation}
Here we have defined the fermionic operators $\hat{d}_{\sigma}^{\dag}$ ($\hat{d}_{\sigma}$), which create (annihilate) electrons with spin $\sigma$ on the QD,  and the corresponding spin-resolved occupation number operators are $\hat{n}_\sigma = \hat{d}_{\sigma}^{\dag}\hat{d}_{\sigma}$, $\sigma=\uparrow, \downarrow$. The operators $\hat{c}_{\alpha k \sigma}^{\dag}$ and $\hat{c}_{\alpha k \sigma}$ create and annihilate, respectively, electrons with momentum $k$, spin $\sigma$, and energy $\varepsilon_{\alpha k \sigma}$ in the source ($\alpha=S$) or drain ($\alpha=D$) electrodes.  We assume that the tunneling matrix elements $t_{\alpha k}$ are independent of spin and consider the situation without an applied magnetic field. Both of these assumptions can be lifted, although such extensions of the model are not considered here.

In the following, we consider weak coupling between the QD and the leads. Electron transport through the QD can then be described by a master equation for transitions between different many-body eigenstates of the QD. The eigenstates of the isolated QD corresponding to Eq.\ (\ref{eq:H_d}) are $|0\rangle$, $|\!\uparrow\rangle$, $|\!\downarrow\rangle$, and $|2\rangle$, where the first and the last eigenstate correspond to the QD being occupied by zero or two electrons, respectively, while the other two correspond to the QD being occupied by a single electron with spin $\sigma=\uparrow,\downarrow$, respectively. We now define the $n$-resolved probabilities\cite{Makhlin2001,Bagrets2003} $p_0(n,t)$, $p_{\uparrow}(n,t)$, $p_{\downarrow}(n,t)$, and $p_2(n,t)$ for each of the eigenstates to be occupied, while $n$ electrons have been collected in the drain during the time span $[0,t]$. Since tunneling is spin independent, it is equally probable to occupy each single-electron spin state, $p_{\downarrow}(n,t)=p_{\uparrow}(n,t)$, and we can define $p_1(n,t)=p_{\downarrow}(n,t)+p_{\uparrow}(n,t)$ and collect the probabilities in the vector \begin{equation}
|p(n,t)\rrangle=[p_0(n,t),p_1(n,t),p_2(n,t)]^T,
\end{equation}
where we use double-brackets to avoid confusion with the quantum states of the Hamiltonian. We also define $\llangle \tilde{0}| = [1,1,1]$, allowing us to express the probability $P(n,t)$ as the inner product $P(n,t)=\llangle\tilde{0}|p(n,t)\rrangle$. The GF is then
\begin{equation}
\mathcal{G}(z,t)=\llangle\tilde{0}|g(z,t)\rrangle,
\end{equation}
where $|g(z,t)\rrangle=\sum_n z^n|p(n,t)\rrangle$. We note that $|g(1,t)\rrangle = [p_0(t),p_1(t),p_2(t)]^T$ contains the probabilities $p_i$ of occupying the QD with $i=0,1,2$ electrons independently of the number of transferred electrons $n$. As a consequence of probability conservation, we have $\mathcal{G}(1,t)=\llangle \tilde{0}|g(1,t)\rrangle=\sum_n P\left(n,t\right)=1$ for all $t$.

The time dependence of the GF is determined by the dynamics of $|g(z,t)\rrangle$. We find the time evolution of $|g(z,t)\rrangle$ by setting up a master equation for the transitions between the eigenstates of the QD. In the weak coupling regime, the transition rates can be found using Fermi's Golden rule, treating the tunneling Hamiltonian in Eq.\ (\ref{eq:H_T}) as the perturbation.\cite{Bruus2004} Working in the wide-band limit and assuming a constant tunneling density of states in both leads, we define the bare energy-independent tunneling rates $\Gamma_\alpha=2\pi|t_{k\alpha}|^2\mathcal{D}_\alpha$, where $\mathcal{D}_\alpha$ is the density of states in lead $\alpha=S,D$. The master equation for $|g(z,t)\rrangle$ then reads
\begin{equation}
\partial_t|g(z,t)\rrangle=\mathbf{M}(z)|g(z,t)\rrangle,
\label{eq:rate_eq}
\end{equation}
with the $z$-dependent rate matrix
\begin{widetext}
\begin{equation*}
\mathbf{M}(z)=  \\ \left[\!\!
                        \begin{array}{ccc}
                          -2(\Gamma_Sn_{S}^{(0)}+\Gamma_Dn_{D}^{(0)})& z\Gamma_D(1-n_{D}^{(0)})+\Gamma_S(1-n_{S}^{(0)}) & 0 \\
                          2(\Gamma_Sn_{S}^{(0)}+z^{-1}\Gamma_Dn_{D}^{(0)}) & -\Gamma_D(1+n_{D}^{(U)}-n_{D}^{(0)})-\Gamma_S(1+n_{S}^{(U)}-n_{S}^{(0)}) & 2\{\Gamma_S(1-n_{S}^{(U)})+z\Gamma_D(1-n_{D}^{(U)})\} \\
                          0 &  z^{-1}\Gamma_Dn_{D}^{(U)}+\Gamma_Sn_{S}^{(U)} &  -2\{\Gamma_S(1-n_{S}^{(U)})+\Gamma_D(1-n_{D}^{(U)})\} \\
                        \end{array}
                      \!\!\right].
\end{equation*}
\end{widetext}
Here, we have introduced the Fermi functions of the leads, kept at electron temperature $T_{\alpha}$ and chemical potential $\mu_\alpha$, evaluated at the energy $\varepsilon_d+E$
\begin{equation}
n_{\alpha}^{(E)}=\frac{1}{e^{(\varepsilon_d+E-\mu_\alpha)/k_BT_\alpha}+1}, \alpha=S,D.
\end{equation}
In the rate matrix above, only $n_{\alpha}^{(0)}$ and $n_{\alpha}^{(U)}$ appear, which we use to parametrize the applied voltage biases and the temperatures of the electrodes in the following subsection. We note that the off-diagonal elements of $\mathbf{M}(z)$ include factors of $z$ and $z^{-1}$ multiplying the rates corresponding to processes that increase or decrease, respectively, by 1 the number of electrons that have been collected in the drain.\cite{Bagrets2003}

We find the GF by formally solving Eq.\ (\ref{eq:rate_eq}) for $|g(z,t)\rrangle$. To this end, we need to define an appropriate initial condition. We assume that the occupations of the QD eigenstates have reached their steady state at $t=0$ when counting begins and $P(n,t=0)=\delta_{n,0}$. We then have $|g(z,t=0)\rrangle=|0\rrangle$, where the stationary state $|0\rrangle$ of the QD is given by the unique solution to $\mathbf{M}(z=1)|0\rrangle=0$. We thereby obtain the following compact expression for the GF:
\begin{equation}
\mathcal{G}(z,t)=\llangle\tilde{0}|e^{\mathbf{M}(z)t}|0\rrangle.
\label{eq:GF_sol}
\end{equation}
This is a general and formally exact result, but in practice, given a rate matrix $\mathbf{M}(z)$, it may not be possible to obtain a simple, closed-form expression for the GF. Further complications arise when trying to calculate (factorial) moments and (factorial) cumulants, since derivatives of the GF with respect to $z$ are required. To calculate high-order (factorial) cumulants at finite times we have thus developed the method described in Appendix~\ref{app:finite_time_fcs}.

\begin{figure*}
\begin{center}
\includegraphics[width=0.95\textwidth]{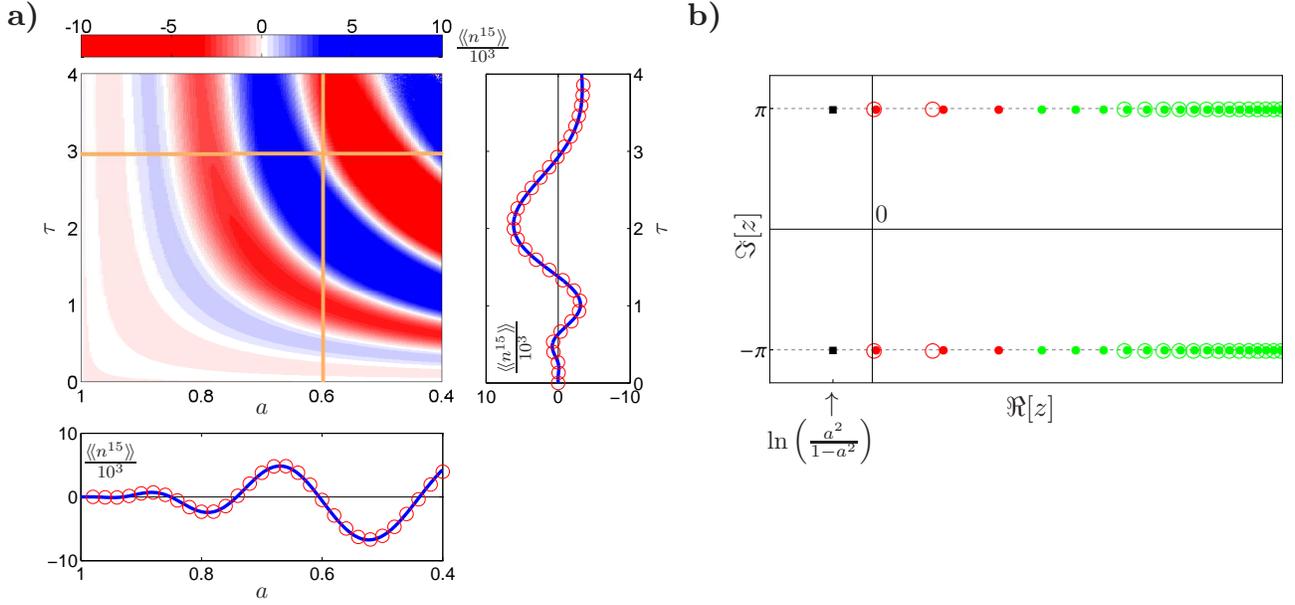}
\caption[]{(Color online) High-order cumulants in the low-bias regime, $n_S^{(U)}\simeq 0$. (a) The cumulant $\llangle n^{15}\rrangle$ as a function of the asymmetry $a$ [Eq.\ (\ref{eq:asymmetry})] and the dimensionless time $\tau=2\Gamma_S t$. The high-order cumulant oscillates as a function of both parameters, in agreement with the experimental results of Fricke \emph{et al.} in Ref.\ \onlinecite{Fricke2010b}. (Note: We do not normalize our results by $\llangle n\rrangle$ as in Ref.\ \onlinecite{Fricke2010b}.) The figures below and next to the contour plot show results along the orange lines $\tau =3 $ and $a=0.6$ in the contour plot. Solid lines are numerical results, while open circles correspond to approximation (\ref{eq:high_cumu}) taking into account the three pairs of singularities indicated by filled red circles in (b).  (b) Singularities of the CGF in the complex plane for  $\tau=3$ and $a=0.6$. The singularities are $2\pi$-periodic along the imaginary axis. Filled circles indicate numerical results obtained from Eq.\ (\ref{Eq:Trans_Eq}), while open red and green circles show approximations (\ref{Eq:Trans_Eq_sol1}) and (\ref{Eq:Trans_Eq_sol2}), respectively. Filled red circles indicate the three pairs of singularities entering Eq.\ (\ref{eq:high_cumu}), as shown by open circles in (a). With time the singularities move toward the points indicated by filled (black) squares.}
\label{Fig:APL}
\end{center}
\end{figure*}

\subsection{Results}

We concentrate on the situation, where the leads are voltage biased such that the energy level of the QD is well above the chemical potential of the drain, $\varepsilon_d\gg\mu_D$ or $n_D^{(0)}=n_D^{(U)}\simeq0$, and electrons cannot tunnel back into the QD from the drain. At the same time, the level is below the chemical potential of the source electrode, $\varepsilon_d\ll\mu_S$ or $n_S^{(0)}\simeq1$. Under these voltage conditions, electron transport takes place from source to drain via the QD. The singly occupied state of the QD always participates in transport, but the doubly occupied state only becomes populated for nonzero values of $n_S^{(U)}$. In the following we study the charge transport statistics as a function of $n_S^{(U)}$, that is, we vary the bias in the source electrode such that $n_S^{(U)}$ takes values between $0$ and $1$.

\subsubsection{Low bias}

We first analyze the low-bias regime $n_S^{(U)}\simeq 0$. This corresponds to the situation recently investigated in the experiments described in Refs.\ \onlinecite{Flindt2009,Fricke2010a,Fricke2010b}. In Ref.\ \onlinecite{Fricke2010b} it was found experimentally that the high-order cumulants oscillated as functions of the dimensionless time
\begin{equation}
\tau=2\Gamma_S t
\label{eq:dimensionlesst}
\end{equation}
and the asymmetry
\begin{equation}
a=\frac{2\Gamma_S-\Gamma_D}{2\Gamma_S+\Gamma_D}.
\label{eq:asymmetry}
\end{equation}
Here, we have included a factor of 2, corresponding to the two spin species in our model. In Fig.\ \ref{Fig:APL}a we show numerically exact results for the high-order cumulant $\llangle n^{15}\rrangle$ corresponding to the measurements in Ref.\ \onlinecite{Fricke2010b}. Our calculations agree well with the experiment and reproduce the clear oscillations as functions of the dimensionless time $\tau$ and the asymmetry $a$. To understand the experimental and numerical results we next consider the zeros of the GF.

At low voltages, the doubly occupied state of the QD remains unpopulated in the stationary state and we can set $p_2(n,t)=0$. This implies a significant simplification of Eq.\ (\ref{eq:rate_eq}), which in this case can be written as
\begin{equation}
\frac{\partial}{\partial t}\left[\!\begin{array}{c}
  g_0(z,t) \\
  g_1(z,t)
\end{array}\!\right]= \left[\!
                     \begin{array}{cc}
                       -2\Gamma_S &  z\Gamma_D\\
                        2\Gamma_S & -\Gamma_D\\
                     \end{array}
                   \!\right] \left[\!
                                  \begin{array}{c}
                                     g_0(z,t) \\
                                     g_1(z,t)
                                  \end{array}
                                \!\right],
\label{Eq:MasterEq2state}
\end{equation}
where $g_i(z,t)=\sum_n z^n p_i(n,t)$, $i=0,1$. We can now explicitly evaluate the GF  and we obtain the analytic function
\begin{equation}
\begin{split}
\mathcal{G}(z,\tau)=\frac{e^{\frac{-\tau}{1+a}}}{4\eta(z)}\left\{e^{\frac{\eta(z)\tau}{1+a} }[1+\eta(z)]^2-e^{-\frac{\eta(z)\tau}{1+a}}[1-\eta(z)]^2\right\}
\end{split}
\label{eq:GF_2state}
\end{equation}
having introduced
\begin{equation}
\eta(z)=\sqrt{a^2+z(1-a^2)}.
\label{eq:eta_func}
\end{equation}
In Eq. (\ref{eq:GF_2state}) we have corrected for a missing factor of $4\eta(z)$ in the Supporting Information in Ref.\ \onlinecite{Flindt2009}. This factor ensures that the GF is an analytic function of $z$, since $-\eta(z)$ is the analytic continuation of $\eta(z)$ across the branch cut and $\mathcal{G}(z,\tau)$ is invariant under the substitution $\eta \rightarrow -\eta$. Solving next for the zeros $z_k$ of the GF, we find
\begin{equation}
z_k=-\frac{h_k^2+a^2}{1-a^2}, k=1,2,\ldots,
\label{eq:GF_zeros}
\end{equation}
where $h_k$ is determined by the transcendental equation
\begin{equation}
h_k\tau=(1+a)\left(k\pi-2\arctan h_k\right),  k=1,2,\ldots.
\label{Eq:Trans_Eq}
\end{equation}
By graphical inspection of the transcendental equation we find that the solutions $h_k$ are real and positive for all $k=1,2,\ldots$. Therefore, all zeros [Eq.\ (\ref{eq:GF_zeros})] are real and negative, since $0\leq a^2\leq 1$. Remarkably, the positions of the zeros agree with the general statements\cite{Abanov2008,Abanov2009} by Abanov and Ivanov for \emph{noninteracting} electrons. This can be understood by noting that the master equation [Eq.\ (\ref{Eq:MasterEq2state})] can also describe noninteracting spinless electrons tunneling through a resonant level for which the statements by Abanov and Ivanov directly apply.

\begin{figure*}
\begin{center}
\includegraphics[width=0.92\textwidth]{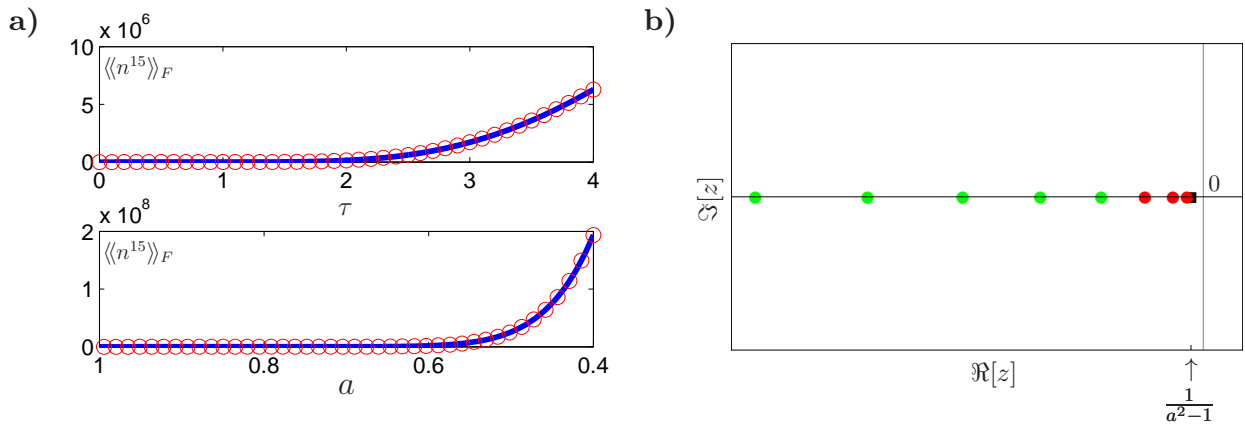}
\caption{(Color online) High-order factorial cumulants in the low-bias regime, $n_S^{(U)}\simeq 0$. (a) The factorial cumulant $\llangle n^{15}\rrangle_F$ as a function of the dimensionless time $\tau=2\Gamma_S t$ and the asymmetry $a$ [Eq.\ (\ref{eq:asymmetry})]. Parameters are $a=0.6$ (upper plot) and $\tau=3$ (lower plot), corresponding to the orange lines in Fig.\ \ref{Fig:APL}(a). Solid lines are numerical results, while open circles correspond to approximation (\ref{eq:high_cumu}) taking into account the three singularities indicated by filled red circles in (b). (b) Singularities of the FCGF in the complex plane for  $\tau=3$ and $a=0.6$. Filled red circles indicate the three singularities entering Eq.\ (\ref{eq:high_cumu}) as shown by open circles in (a). With time the singularities move toward the point indicated by the filled (black) square. Singularities lie on the negative real axis, and the factorial cumulants do not oscillate as functions of either time or asymmetry.}
\label{Fig:APLfc}
\end{center}
\end{figure*}

In the following, we solve the transcendental equation (\ref{Eq:Trans_Eq}) numerically. However, in two limiting cases, it can be solved analytically: For any fixed $k$, we can choose a sufficiently large $\tau$, such that $h_{k}$ must be small and $\arctan h_{k}\simeq h_{k}$. We then obtain
\begin{equation}
h_{k}\simeq \left(\frac{1+a}{1+a+\frac{\tau}{2}}\right)\frac{\pi}{2}k, \,\,\,k\,\,\, \mathrm{fixed}, \,\,\,\tau\,\,\,  \mathrm{large}.
\label{Eq:Trans_Eq_sol1}
\end{equation}
In the other limiting case, we fix the time $\tau$ and consider a large $k$, such that $h_{k}$ must be large and $\arctan h_{k}\simeq\pi/2$. We then find
\begin{equation}
h_{k}\simeq \left(\frac{1+a}{\tau}\right)\pi(k-1), \,\,\,\tau\,\,\, \mathrm{fixed}, \,\,\,k\,\,\,  \mathrm{large}.
\label{Eq:Trans_Eq_sol2}
\end{equation}
From the zeros of the GF we can understand the behavior of high-order (factorial) cumulants. Analyzing first the ordinary cumulants, we note that the CGF has logarithmic singularities at $\ln |z_k|+i\pi(2l+1)$, $l\in \mathbb{Z}$, corresponding to the zeros $z_k$ of the GF. In Fig.\ \ref{Fig:APL}b we show the positions of these singularities obtained from numerical solutions of the transcendental equation [Eq.\ (\ref{Eq:Trans_Eq})]. For comparison, we also show the limiting cases, Eqs.\ (\ref{Eq:Trans_Eq_sol1}) and (\ref{Eq:Trans_Eq_sol2}), which show good agreement with the numerical results. In Fig.\ \ref{Fig:APL}a, we show numerically exact results together with the asymptotic expression [Eq.\ (\ref{eq:high_cumu})], taking into account the three pairs of complex-conjugate singularities that are closest to $z=0$ [indicated by filled red circles in Fig.\ \ref{Fig:APL}(b)]. The agreement is good and the analysis provides us with a simple interpretation of the experimental data from Ref.\ \onlinecite{Fricke2010b}: The motion of the singularities in the complex plane as functions of the dimensionless time $\tau$ and the asymmetry $a$ cause the oscillations of the high-order cumulants observed in our numerical calculations and in the experiment.

We next turn to the corresponding factorial cumulants. Numerical results for the factorial cumulant of order $m=15$ as a function of the dimensionless time $\tau$ and the asymmetry $a$ are shown in Fig.\ \ref{Fig:APLfc}a and clearly no oscillatory behavior is observed. Again, this can be understood by considering the logarithmic singularities of the FCGF. The FCGF has logarithmic singularities at $z_k-1$ corresponding to the (negative) zeros $z_k$ of the GF. Since the singularities of the FCGF are real, the factorial cumulants do not oscillate as functions of either $\tau$ or $a$ [according to Eq.\ (\ref{eq:high_cumu_onesing})]. In Fig.\  \ref{Fig:APLfc}b we show the positions of the singularities, and in Fig.\  \ref{Fig:APLfc}a we show the asymptotic expression [Eq.\ (\ref{eq:high_cumu})], taking into account the three singularities that are closest to $z=0$ [indicated by filled red circles in Fig.\ \ref{Fig:APLfc}(b)]. The asymptotic expression fully accounts for the numerical results.

Before closing this part, we briefly discuss the motion of the singularities at long times. As indicated in Figs.\ \ref{Fig:APL}b and \ref{Fig:APLfc}b,  the singularities of the CGF and the FCGF all move toward the points marked by filled black squares. According to Eq.\ (\ref{eq:GF_zeros}), the zeros of the GF behave as $z_k\rightarrow a^2/(a^2-1)\leq 0$ for long times $\tau$, since then $h_k\simeq 0$ in Eq.\ (\ref{Eq:Trans_Eq_sol1}). The points marked by filled black squares in Figs.\ \ref{Fig:APL}b and \ref{Fig:APLfc}b are thus $\ln [a^2/(a^2-1)]$ (for the CGF) and $a^2/(a^2-1)-1=1/(a^2-1)$ (for the FCGF), respectively. Interestingly, the point $z=a^2/(a^2-1)$ corresponds to the square-root branch point of the function $\eta(z)$ in Eq.\ (\ref{eq:eta_func}): In the long-time limit, the logarithm of the GF goes as $\ln \mathcal{G}(z,\tau)\rightarrow \tau[\eta(z)-1]/(1+a)$ according to Eq.\ (\ref{eq:GF_2state}), and the singularities of the CGF and the FCGF are then determined by the branch points of $\eta(z)$. Thus, in this example, all logarithmic singularities of the CGF and the FCGF move toward particular points in the complex plane, which in the long-time limit become branch point singularities.

\subsubsection{Finite bias}

In the general case of a finite bias, $0< n^{(U)}_S< 1$, the on-site Coulomb interaction strongly influences the charge transport statistics and we expect that the statistics will no longer be generalized binomial, with all zeros of the GF lying on the negative real axis. It is now a difficult task to write down a closed-form analytic expression for the GF and its zeros. However, as we will see, the positions of the dominant zeros in the complex plane can be deduced from the high-order factorial cumulants.

\begin{figure}
\begin{center}
\includegraphics[width=0.38\textwidth]{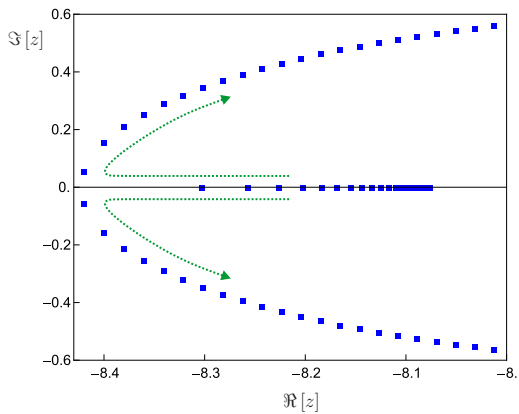}
\caption{(Color online) Motion of the dominant singularities of the FCGF $\ln \left[\mathcal{G}(z+1,t)\right]$ into the complex plane. The asymmetry is $a=0.89$ and the dimensionless time is $\tau=2\Gamma_S t=4$. Dotted arrows indicate the motion of the zeros as $n_S^{(U)}$ is increased from 0.75 to 0.85 in steps of 0.0025. Singularities start off on the negative real axis, but eventually they move into the complex plane.}
\label{Fig4}
\end{center}
\end{figure}

We first demonstrate how the zeros of the GF move into the complex plane due to interactions. This is illustrated in Fig.\ \ref{Fig4}, where we show numerical results for the logarithmic singularities of the FCGF as functions of $n^{(U)}_S$. We have solved numerically the equation $\mathcal{G}(z+1,t)=0$ and show in Fig.\ \ref{Fig4} the singularities that are closest to $z=0$. For low biases, $n^{(U)}_S\simeq 0$, the dominant singularity is on the negative real axis, but as the bias increases, it is eventually reached by the second-closest singularity at a degeneracy point from which the pair of singularities splits off from the real axis and moves into the complex plane. At this point, the statistics is no longer generalized binomial due to the on-site Coulomb interaction.

\begin{figure}
\begin{center}
\includegraphics[width=0.45\textwidth]{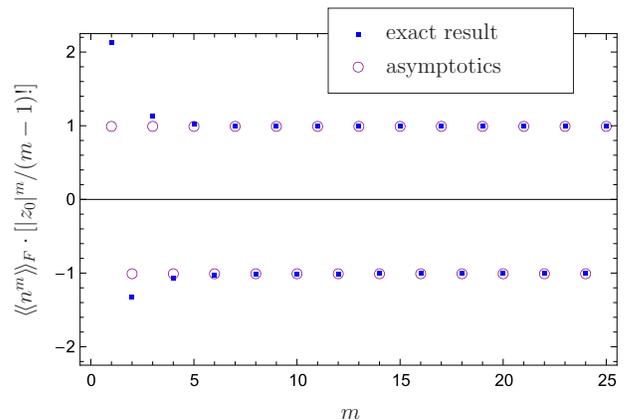}
\caption{(Color online) Factorial cumulants $\llangle n^m \rrangle_F$ as functions of the order $m$. Parameters are $a=0.89$, $\tau=2\Gamma_S t= 4$, and $n_S^{(U)}=0.4$, for which the dominant singularity $z_0\simeq -8.2$ of the FCGF $\ln\left[\mathcal{G}(z+1,t)\right]$ is real and negative. Factorial cumulants have been rescaled as $\llangle n^m \rrangle_F\cdot[|z_0|^m/(m-1)!]$. Numerical results (filled squares) follow the predicted behavior for high-order factorial cumulants given by Eq.\ (\ref{eq:high_cumu_onesing}) (open circles).}
\label{Fig5}
\end{center}
\end{figure}

The change of statistics is directly reflected in the high-order factorial cumulants. In Fig.\ \ref{Fig5} we show the factorial cumulants $\llangle n^m \rrangle_F$ as functions of their order $m$ with parameters for which the dominant singularity is still real and negative, such that the high-order factorial cumulants should follow Eq.\ (\ref{eq:high_cumu_onesing}). In Fig.\ \ref{Fig5} we have rescaled the factorial cumulants as $\llangle n^m \rrangle_F\cdot[|z_0|^m/(m-1)!]$, where $z_0$ is the singularity closest to $z=0$, and for high orders they clearly follow the predicted pattern with alternating signs due to the factor $(-1)^{(m-1)}$ in Eq.\ (\ref{eq:high_cumu_onesing}). The asymptotic behavior of the rescaled high-order factorial cumulants (shown by open circles) agrees well with our numerical results (filled squares). From the fifth factorial cumulant and onward, the asymptotic expression completely accounts for our numerical results.

\begin{figure}
\begin{center}
\includegraphics[width=0.45\textwidth]{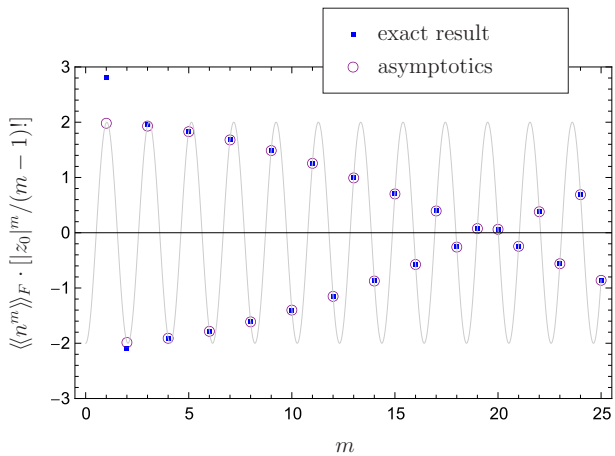}
\caption{(Color online) Factorial cumulants $\llangle n^m \rrangle_F$ as functions of the order $m$. Parameters are $a=0.89$, $\tau=2\Gamma_S t= 4$, and $n_S^{(U)}=0.9$, for which the high-order factorial cumulants are dominated by the pair of complex-conjugate singularities $z_0\simeq -7.6+0.62\,i$ and $z_0^*$. Factorial cumulants have been rescaled as $\llangle n^m \rrangle_F\cdot[|z_0|^m/(m-1)!]$. Numerical results (filled squares) follow the predicted behavior for the high-order factorial cumulants given by Eq.\ (\ref{eq:high_cumu_twosing}) (open circles). The solid line serves as a guide for the eye.}
\label{Fig6}
\end{center}
\end{figure}

In Fig.\ \ref{Fig6} we show results for the factorial cumulants with parameters for which the dominant singularities have moved into the complex plane and the statistics is no longer generalized binomial. The high-order factorial cumulants are now governed by a pair of complex-conjugate singularities and are thus expected to follow Eq.\ (\ref{eq:high_cumu_twosing}). In Fig.\ \ref{Fig6} we have rescaled the factorial cumulants as $\llangle n^m \rrangle_F\cdot[|z_0|^m/(m-1)!]$, where $z_0$ and $z_0^*$ are the closest complex-conjugate pair of singularities. In this case, we see clear trigonometric oscillations as a function of the order $m$ due to the factor $\cos{(m\arg [z_0])}$ with a frequency determined by $\arg [z_0]$. The asymptotic expression (open circles) accounts well for our numerical results for the factorial cumulants (filled squares) already from the third order onward. We stress that this oscillatory behavior of the factorial cumulants would not be possible in a noninteracting system for which all singularities would lie on the negative real axis such that $\arg [z_0]=\pi$ and $\cos{(m\arg [z_0])}=(-1)^{(m-1)}$.

\begin{figure}
\begin{center}
\includegraphics[width=0.44\textwidth]{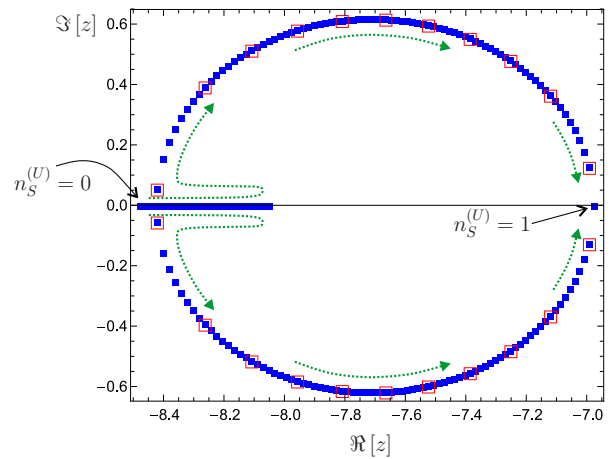}
\caption{(Color online) Motion of the dominant singularities of the FCGF in the complex plane as a function of the parameter $n_S^{(U)}$. The asymmetry is $a=0.89$ and the dimensionless time is $\tau=2\Gamma_S t=4$. Arrows indicate how the singularities move as $n_S^{(U)}$ is increased from 0 to 1 in steps of 0.0025. Numerical solutions of the singularities are shown by filled squares, while open squares are the positions obtained from the high-order factorial cumulants (up to order $m=25$) as explained in Appendix \ref{app:B}.}
\label{Fig7}
\end{center}
\end{figure}

For the results presented in Figs.\ \ref{Fig4} to \ref{Fig6} we found numerically the logarithmic singularities of the FCGF. In an actual experiment, however, the GF would typically not be known, and it is therefore relevant to ask if the position of the singularities can be deduced from the measured high-order factorial cumulants alone. In Appendix \ref{app:B} we describe a simple method for extracting the position of a dominant pair of complex-conjugate singularities from four consecutive high-order factorial cumulants. As illustrated in Fig.\ \ref{Fig7}, we find excellent agreement between our numerical results and the positions obtained directly from the high-order cumulants (up to order $m=25$) using this method. The figure shows results in the full range of bias voltages from $n_S^{(U)}=0$ to $n_S^{(U)}=1$. For low bias voltages $n_S^{(U)}\simeq0$, the dominant singularity is real and negative, and as $n_S^{(U)}$ is increased, two complex-conjugate singularities move into the complex plane, showing that the statistics is no longer generalized binomial. Finally, in the high-bias limit $n_S^{(U)}=1$, the singularities move back onto the real axis, and the statistics is again generalized binomial. In the high-bias regime, the interaction strength $U$ drops out of the problem and transport takes place via the two parallel and uncorrelated spin channels that independently give rise to generalized binomial statistics.

\section{Conclusions}
\label{sec:conclusions}

We have shown that factorial cumulants are useful for detecting interactions among electrons passing through a nanoscale device. For noninteracting electrons in a two-terminal conductor, the counting statistics is always generalized binomial, as recently found by Abanov and Ivanov:\cite{Abanov2008,Abanov2009} The charge transport statistics can be factorized into single-particle transfer events and the zeros of the GF are correspondingly real and negative. Interactions among the electrons, however, can drive the zeros of the GF away from the negative real axis and into the complex plane.\cite{Abanov2009} As we have shown, this change of the statistics is clearly visible in the factorial cumulants. For generalized binomial statistics, the factorial cumulants have a sign that is determined by the cumulant order only. In contrast, as the zeros of the GF move into the complex plane due to interactions, the factorial cumulants oscillate as functions of basically any parameter.

To illustrate our findings, we have considered transport through a quantum dot weakly coupled to source and drain electrodes. At low bias voltages, the dot can only be empty or singly occupied. This corresponds to recent experiments\cite{Fricke2010b} for which we reproduce and explain the measured oscillations of the high-order cumulants by considering the zeros of the GF. In this case, the statistics is generalized binomial and the high-order \emph{factorial} cumulants do not oscillate. As the bias voltage is increased an additional electron can occupy the dot. The on-site interaction now strongly affects the counting statistics, and the zeros of the GF move into the complex plane. As we have shown, this is clearly visible in the high-order factorial cumulants, which also allow us to locate the positions of the zeros. We expect that the motion of zeros into the complex plane due to interactions will be experimentally detectable using available measurement techniques.

Our work leaves a number of open questions for future research. It would be interesting to understand physically the exact point at which interactions cause the zeros of the GF to become complex. We have considered only two-terminal devices, and it might be possible to generalize our findings to multilead setups.\cite{Kambly2009}

\begin{appendix}

\acknowledgements

We thank D.\ A.\ Ivanov for illuminating discussions, C.\ Fricke and N.\ Sethubalasubramanian for useful correspondences about their experiment, and C.\ Bruder and P.\ H\"anggi for drawing our attention to factorial cumulants. This work is supported by the Swiss NSF, the Swiss center for excellence MaNEP, the European Network NanoCTM, and the Carlsberg Foundation.

\section{Calculation of (factorial) cumulants at finite times}
\label{app:finite_time_fcs}

Here we describe our method for calculating (factorial) cumulants at finite times without explicitly evaluating the (F)CGF.
We begin with a master equation of the form given by Eq.\ (\ref{eq:rate_eq}),
\begin{equation}
\partial_t|g(z,t)\rrangle=\mathbf{M}(z)|g(z,t)\rrangle,
\end{equation}
for a system with $N$ states. Next, we substitute in the master equation $|g(z,t)\rrangle$ by $|\tilde{g}(z,t)\rrangle$ and $\mathbf{M}(z)$ by $\widetilde{\mathbf{M}}(z)$, where $|\tilde{g}(z,t)\rrangle=|g(e^z,t)\rrangle$ and $\widetilde{\mathbf{M}}(z)=\mathbf{M}(e^z)$ for ordinary cumulants, and $|\tilde{g}(z,t)\rrangle=|g(z+1,t)\rrangle$ and $\widetilde{\mathbf{M}}(z)=\mathbf{M}(z+1)$ for factorial cumulants, respectively. For $z=0$, the master equation then reads
\begin{equation}
\partial_t|\tilde{g}^{(0)}(t)\rrangle = \widetilde{\mathbf{M}}^{(0)}|\tilde{g}^{(0)}(t)\rrangle,
\label{eq:rate_eq_z=0}
\end{equation}
having defined $|\tilde{g}^{(0)}(t)\rrangle=|\tilde{g}(0,t)\rrangle$ and $\widetilde{\mathbf{M}}^{(0)}=\widetilde{\mathbf{M}}(0)$. Taking instead $m=1,2,\ldots, k$ consecutive derivatives with respect to $z$  (evaluated at $z=0$), we obtain
\begin{eqnarray}
\partial_t|\tilde{g}^{(1)}(t)\rrangle &=& \widetilde{\mathbf{M}}^{(1)}|\tilde{g}^{(0)}(t)\rrangle+ \widetilde{\mathbf{M}}^{(0)}|\tilde{g}^{(1)}(t)\rrangle, \nonumber \\
\partial_t|\tilde{g}^{(2)}(t)\rrangle &=&  \widetilde{\mathbf{M}}^{(2)}|\tilde{g}^{(0)}(t)\rrangle+2 \widetilde{\mathbf{M}}^{(1)}|\tilde{g}^{(1)}(t)\rrangle \nonumber\\
&&+ \widetilde{\mathbf{M}}^{(0)}|\tilde{g}^{(2)}(t)\rrangle,\nonumber\\
&\vdots& \nonumber\\
\partial_t|\tilde{g}^{(k)}(t)\rrangle &=& \sum_{j=0}^k \binom{k}{j} \widetilde{\mathbf{M}}^{(j)}|\tilde{g}^{(k-j)}(t)\rrangle,
\label{eq:coupled_system}
\end{eqnarray}
where we have introduced the notation $|\tilde{g}^{(m)}(t)\rrangle=\partial_z^m|\tilde{g}(z,t)\rrangle|_{z\rightarrow 0}$  and $\widetilde{\mathbf{M}}^{(m)}=\partial_z^m\widetilde{\mathbf{M}}(z)|_{z\rightarrow 0}$. The (factorial) moments of order $m\leq k$ are then
\begin{equation}
\langle n^m\rangle_{(F)}(t)=\llangle\tilde{0}|\tilde{g}^{(m)}(t)\rrangle,
\label{eq:cumu_numeric}
\end{equation}
depending on the substitutions made above. The vector $\llangle\tilde{0}|=[1,1, \ldots, 1]$ contains $N$ elements equal to unity.

We solve the system of coupled equations (\ref{eq:rate_eq_z=0},\ref{eq:coupled_system}) by introducing the auxiliary vector
\begin{equation}
|G(t)\rrangle=[|\tilde{g}(t)\rrangle,|\tilde{g}^{(1)}(t)\rrangle,\ldots,|\tilde{g}^{(k)}(t)\rrangle]^T
\end{equation}
containing $N(k+1)$ elements. The equation of motion for $|G(t)\rrangle$ reads
\begin{equation}
\partial_t|G(t)\rrangle=\mathbf{\underline{\underline{M}}}|G(t)\rrangle
\label{eq:fullEOM}
\end{equation}
where according to Eqs.\ (\ref{eq:rate_eq_z=0},\ref{eq:coupled_system})
\begin{equation}
\mathbf{\underline{\underline{M}}}=
            \left[\begin{array}{ccccc}
                            \widetilde{\mathbf{M}}^{(0)} & 0 & 0 & 0 & 0 \\
                            \widetilde{\mathbf{M}}^{(1)} & \widetilde{\mathbf{M}}^{(0)} & 0 & 0 & 0 \\
                            \widetilde{\mathbf{M}}^{(2)} & 2\widetilde{\mathbf{M}}^{(1)} & \widetilde{\mathbf{M}}^{(0)} & 0 & 0 \\
                            \vdots & \vdots  & ~ & \ddots & ~ \\
                            \widetilde{\mathbf{M}}^{(k)} & ~ & ~ & n\widetilde{\mathbf{M}}^{(1)} & \widetilde{\mathbf{M}}^{(0)} \\
            \end{array}\right].
            \label{eq:Mtotal}
\end{equation}
is a matrix of dimensions $N(k+1)\times N(k+1)$. We proceed by solving Eq.\ (\ref{eq:fullEOM}) as
\begin{equation}
|G(t)\rrangle=e^{\mathbf{\underline{\underline{M}}}t}|G(t=0)\rrangle.
  \label{eq:EOMsol}
\end{equation}
Here, the initial condition as counting begins reads
\begin{equation}\label{SolBigMat}
    |G(t=0)\rrangle=[|0\rrangle,0,0,\ldots,0]^T,
\end{equation}
and contains the stationary state $|0\rrangle$, which solves $\widetilde{\mathbf{M}}^{(0)}|0\rrangle=0$, followed by $N k$ elements equal to 0. Even for large dimensions of $\mathbf{\underline{\underline{M}}}$ we may calculate numerically the matrix exponentiation $e^{\mathbf{\underline{\underline{M}}}t}$ for a given time $t$ and obtain $|G(t)\rrangle$ via Eq.\ (\ref{eq:EOMsol}). Having determined the (factorial) moments using Eq.\ (\ref{eq:cumu_numeric}), the corresponding (factorial) cumulants are obtained via the relation\cite{Johnson2005}
\begin{equation}
    \llangle n^m\rrangle_{(F)}=\langle n^m\rangle_{(F)}-\sum_{k=1}^{m-1}\binom{m-1}{k-1}\llangle n^k\rrangle_{(F)}\bigskip\langle n^{m-k}\rangle_{(F)}.
\label{eq:Moments2Cumulants}
\end{equation}
For the particular $N=3$ state model studied in this work, we could easily calculate the first $m=50$ (factorial) cumulants at finite times.
\begin{widetext}
\section{Determination of a pair of complex-conjugate singularities}
\label{app:B}
A pair of dominant, complex-conjugate logarithmic singularities $z_0$ and $z_0^*$ can be extracted from four consecutive high-order factorial cumulants using methods from Refs. \onlinecite{Zamastil2005,Flindt2010}, which we repeat here for completeness. Using Eq.~(\ref{eq:high_cumu_twosing}) we obtain the matrix equation
\begin{equation}
\left[
      \begin{array}{cc}
        1 \quad & -\frac{1}{m-2}\frac{\llangle n^{m-1} \rrangle_{(F)}}{\llangle n^{m-2} \rrangle_{(F)}} \\
        1 \quad & -\frac{1}{m-1}\frac{\llangle n^{m} \rrangle_{(F)}}{\llangle n^{m-1} \rrangle_{(F)}} \\
      \end{array}
    \right]\cdot
    \left[
      \begin{array}{c}
        2\Re\left(z_0\right) \\
        \left|z_0\right|^2 \\
      \end{array}
    \right]=
    \left[
      \begin{array}{c}
        \left(m-3\right)\frac{\llangle n^{m-3} \rrangle_{(F)}}{\llangle n^{m-2} \rrangle_{(F)}} \\
        \left(m-2\right)\frac{\llangle n^{m-2} \rrangle_{(F)}}{\llangle n^{m-1} \rrangle_{(F)}} \\
      \end{array}
    \right],
\end{equation}
which we solve for $\Re\left(z_0\right)$ and $\left|z_0\right|^2$. Typically the accuracy of the method increases with the cumulant order $m$.\cite{Zamastil2005}
\end{widetext}
\end{appendix}


\end{document}